\documentclass[aps,nofootinbib,preprint]{revtex4}
\usepackage{amsmath}
\usepackage{graphicx}

\begin{document}

\title{Renormalization-group analysis for
low-energy scattering\newline of charged particles}
\author{Shung-ichi Ando}\email{shung-ichi.ando@manchester.ac.uk} 
\author{Michael C. Birse}\email{mike.birse@manchester.ac.uk}
\affiliation{
Theoretical Physics Group, School of Physics and Astronomy, 
The University of Manchester, Manchester, M13 9PL, UK}

\begin{abstract}

The low-energy scattering of two charged particles is analyzed 
using a renormalization group approach based on 
dimensional regularization with power-divergence subtraction. 
A nontrivial solution with a marginally unstable direction is
found, corresponding to a system with a bound state at zero
energy. For purely energy-dependent perturbations around 
this solution, the power counting agrees with that from 
Wilsonian methods. These terms in the effective potential 
are in direct correspondence with the the terms in the 
Coulomb-distorted effective-range expansion. We also study 
perturbations that depend on off-shell momenta as well as 
energy, and we show that these affect only the off-shell 
form of the scattering matrix. These terms are of higher 
order that the corresponding energy-dependent ones and so 
terms in the potential that depend only on the off-shell 
momenta do not have definite orders in power counting.
\end{abstract}
\maketitle

\vskip 10pt

\section{Introduction}

The effective field theories (EFTs) that have been developed 
in recent years are now important tools for analyzing
the scattering of low-energy particles, particularly in
the context of nuclear forces. 
(For reviews, see, e.g., Refs.~\cite{betal-00,bk-arnps02,e-ppnp06}.)
They can provide systematic 
expansions of interactions in powers of low-energy scales.
In addition, they form a field-theoretic framework for 
extending the older effective-range expansions developed by
Bethe and others~\cite{b-pr49}. In the context of charged particle 
scattering, Kong and Ravndal~\cite{kr-plb99,kr-prc01} showed how 
Bethe's  Coulomb-modified effective-range expansion could be 
represented as a pionless EFT.

The results of Kong and Ravndal~\cite{kr-plb99,kr-prc01} at NLO 
showed unexpectedly strong dependence of the scattering length on 
the subtraction scale $\mu$. 
More recently, this approach was extended to NNLO~\cite{ashh-prc07} 
and similar dependence of other coefficients on $\mu$ were observed, 
along with dependencies on the renormalization scheme and the choice 
of momentum structures in the NNLO terms. 
These results contrast with ones obtained from an 
analysis using a Wilsonian renormalization group~\cite{bb-prc03}, 
which showed a one-to-one correspondence between terms in the 
effective potential and the effective-range expansion.

The main difference between the effective short-range potentials
used in those two approaches is that those in 
Refs.~\cite{kr-plb99,kr-prc01,ashh-prc07} depend on the off-shell
relative momenta of the two particles whereas those in 
Ref.~\cite{bb-prc03} depend on energy. Applications of the 
renormalization group (RG)~\cite{wk-pr74} to pure short-range 
interactions~\cite{bmr-plb99} have shown that, 
in systems with bound or virtual states close to threshold, 
momentum- and energy-dependent terms have different scale dependence 
and hence appear at different orders in the power counting.  

To examine whether this differences between these short-range 
potentials is responsible, we have made an RG analysis of
their scaling properties. To facilitate the comparison with
Refs.~\cite{kr-plb99,kr-prc01,ashh-prc07} 
we have used dimensional regularization (DR) 
with the power divergence subtraction (PDS)~\cite{ksw-plb98,ksw-npb98} 
and modified minimal subtraction ($\overline{\rm MS}$) schemes. 
We consider potentials that can depend on energy to all orders but, 
for simplicity, we restrict our analysis to the simplest momentum-dependent 
term, which is of second order in the off-shell momenta.

The paper is organized as follows. In Sec.~II, we apply the RG 
to an energy-dependent short-range interaction in the presence 
of the Coulomb potential. We concentrate on the expansion around
a marginally unstable nontrivial solution that is the analogue
of the fixed point discussed in Ref.~\cite{bmr-plb99}. This
allows us to confirm that DR leads to the same RG flow as found 
in Ref.~\cite{bb-prc03} using a sharp cut-off on the 
distorted waves. Then, in Sec.~III, we find the leading
momentum-dependent perturbation around our nontrivial solution. 
This has a different RG eigenvalue from the corresponding 
energy-dependent term. Its form suggests that it contributes 
only to the off-shell dependence of the scattering matrix,
and we conform this by calculating the on-shell scattering
amplitude in Sec.~IV. Finally, we discuss the implications
of this and other features of our results in Sec.~V.

\section{Scattering by long- and short-range forces}

The two-body potential $V$ between two charged
particles can be separated into two pieces:
\begin{equation}
V = V_C+V_S\, ,
\end{equation}
where $V_C$ is the long-range Coulomb potential and
$V_S$ is the short-range strong interaction. For momenta that
are too low to resolve the range of the short-range interaction, 
we can represent this part of the potential by delta functions
and their derivatives. The RG will then lead us to a systematic
power counting for these terms.

For the sum of the two potentials above, the scattering matrix
$T$ can be split into $T_C$, the pure Coulomb $T$ matrix, and 
$T_{SC}$ which describes scattering between distorted waves of 
the Coulomb potential. If we restrict our attention to $S$-waves, 
the on-shell matrix element of $T_{SC}$ can be related to the 
strong-interaction phase shift, $\delta_{SC}(p)$, by
\begin{equation}
\langle \psi^-_C(p)|T_{SC}|\psi^+_C(p)\rangle=-\,\frac{4\pi}{Mp}\, 
\frac{e^{2i\sigma_0(p)}}{\cot\delta_{SC}(p)-i}\,,
\end{equation}
where $\sigma_0$ is the Coulomb phase shift,  
$|\psi^\pm_C(p)\rangle$ are the in- and out-going Coulomb wave 
functions, and $p=\sqrt{ME}$ is the on-shell 
momentum.\footnote{Here we are considering scattering of two 
identical particles of mass $M$. More generally $M$ should be 
replaced by twice the reduced mass.}

At low energies, the $S$-wave phase shift $\delta_{SC}(p)$ 
can be written in the form of an effective-range expansion
as
\begin{equation}
{\cal C}(\kappa/p)\, p\, \cot\delta_{SC}(p) + 2\kappa 
\mbox{Re}\{H(\kappa/p)\}
= -\frac{1}{a_C} + \frac12 r_0 p^2 + \cdots\, ,
\label{eq;ERE}
\end{equation}
where 
\begin{equation}
{\cal C}(\eta) = \frac{2\pi \eta}{e^{2\pi\eta}-1}\, ,
\qquad
H(\eta) = \psi(i\eta) + \frac{1}{2i\eta}-\ln(i\eta)\, ,
\end{equation}
$\kappa$ is the inverse of the Bohr radius 
(we will give its definition below), 
and $\psi$ is the logarithmic derivative of the $\Gamma$
function. In this expansion, $a_C$ is the scattering length 
and $r_0$ is the effective range. 

This $T$ matrix satisfies the Lippmann-Schwinger (LS) equation
\begin{equation}
T_{SC}(E)=V_S + V_S\, G_C(E)\, T_{SC}(E)\, .
\label{eq;LSeq}
\end{equation}	
Here $G_C=1/(E-H_0-V_C+i\epsilon)$ is the Coulomb Green's function. 
In terms of the Coulomb wave functions, this has the form
\begin{equation}
{G}_C(E) 
= M \int \frac{d^3{\bf q}}{(2\pi)^3}\,
\frac{|\psi^+_C(q)\rangle\langle\psi^+_C(q)|}
{p^2-q^2+i\epsilon}\, .
\end{equation}

\section{RG analysis for energy-dependent potentials}

Before examining momentum-dependent short-range potentials, we first 
consider purely energy-dependent ones. This analysis reproduces 
the results of Ref.~\cite{bb-prc03}, but within the framework of
DR and the PDS scheme. This establishes the notation we shall use for more 
general potentials and it allows us to make direct contact with
the approach of Kong and Ravndal \cite{kr-plb99,kr-prc01}, which was
also used in Ref.~\cite{ashh-prc07}.

In DR, we replace the $3+1$-dimensional Coulomb Green's function by
\begin{equation}
G_C(E,\mu) = M\left(\frac{\mu}{2}\right)^{4-d}\int
\frac{d^{d-1}{\bf q}}{(2\pi)^{d-1}}\,\frac{|\psi^+_C(q)\rangle 
\langle \psi^+_C(q)|}
{p^2 - q^2+i\epsilon} \, .
\end{equation}
The resulting integrals have poles at $d=3$ and 4, corresponding to linear
and logarithmic divergences respectively. We subtract all of these, using 
the $\overline{\rm MS}$ scheme for the poles at $d=4$.

For a simple $\delta$-function interaction, its strength 
$V_S(p,\kappa,\mu)$ can depend on three scales: the on-shell 
momentum $p$, the inverse of the Bohr radius,
\begin{equation}
\kappa = \frac{\alpha M}{2}\, ,
\end{equation}
and the subtraction scale $\mu$. The resulting regularized
LS equation for $T_{SC}$ takes the form
\begin{equation}
T_{SC}(p, \kappa)=V_S(p,\kappa,\mu)
+V_S(p,\kappa,\mu)J^{\overline{\rm MS}}_0(p,\kappa,\mu)
T_{SC}(p, \kappa)\, ,
\end{equation}
where the bubble integral is defined by
\begin{equation}
J_0=\int \frac{d^3{\bf k}}{(2\pi)^3}\,
\frac{d^3{\bf k}'}{(2\pi)^3}\,
\langle {\bf k}|G_C|{\bf k}'\rangle\,.
\end{equation}
In the dimensional regularization scheme used here, 
this becomes \cite{kr-prc01,ashh-prc07}
\begin{equation}
J_0^{\overline{\rm MS}}(p,\kappa,\mu) = 
-\,\frac{M}{4\pi}\,\mu +\frac{\kappa M}{2\pi}\left[
1-\gamma +\ln\left(\frac{\mu}{4\kappa}\right)
\right] - \frac{\kappa M}{2\pi}H(\eta)\, ,
\label{eq;J0}
\end{equation} 
where $\gamma$ is the Euler-Mascheroni constant. 
Note that the Coulomb wave functions at the origin,
$\int {d^3{\bf k}}/{(2\pi)^3}\langle {\bf k}|\psi_C(q)\rangle$,
are finite and so the only divergences here arise from the
integral over ${\bf q}$. With momentum-dependent interactions, 
we encounter the derivatives of these functions. These 
diverge at the origin and so require additional 
regularization, as discussed in the next section.

We now demand that the $T$ matrix be independent of the 
arbitrary scale $\mu$. This condition ensures that physical
scattering observables will not depend on $\mu$.
It leads to a differential equation for the effective 
potential, which is similar to that controlling the
cut-off dependence of the potential in Wilsonian
approaches~\cite{bmr-plb99,bb-prc03}. 
This equation has the form
\begin{equation}
\frac{\partial}{\partial \mu}\,V_S(p,\kappa,\mu)
= -V_S(p,\kappa,\mu)^2\,\frac{\partial}{\partial \mu}\,
J_0^{\overline{\rm MS}}(p,\kappa,\mu)
= -V_S(p,\kappa,\mu)^2\left(-\,\frac{M}{4\pi}
+\frac{M}{2\pi}\frac{\kappa}{\mu}\right)\, . 
\label{eq;RGeqV0}
\end{equation}

This equation can be converted into an RG equation by 
expressing all dimensioned quantities in units of $\mu$. 
We define $\hat{p}=p/\mu$ and $\hat\kappa=\kappa/\mu$
and introduce the rescaled potential
\begin{equation}
\hat{V}_S(\hat{p},\hat{\kappa},\mu)
= \frac{M\mu}{4\pi}\, V_S(\mu\hat{p},\mu\hat{\kappa},\mu)\, .
\end{equation}
In terms of these, we can rewrite Eq.~(\ref{eq;RGeqV0}) as
\begin{equation}
\mu\, \frac{\partial}{\partial \mu}
\left(\frac{1}{\hat{V}_S}\right)
= \hat{p}\,\frac{\partial}{\partial \hat{p}}
\left(\frac{1}{\hat{V}_S}\right)
+ \hat{\kappa}\,\frac{\partial}{\partial \hat{\kappa}}
\left(\frac{1}{\hat{V}_S}\right)
- \frac{1}{\hat{V}_S} - \left(1-2\hat{\kappa}\right)\, .
\label{eq;RGeqinvhatV0}
\end{equation}
Below all production thresholds, the effective potential 
should be an analytic function of the energy and all other 
low-energy scales. We thus require solutions to this 
equation to be analytic in $\hat p^2$ and $\hat\kappa$.

Fixed points of the RG equation play a crucial role, as 
discussed in detail in Ref.~\cite{bmr-plb99}. In the limit 
$\mu\to 0$, the rescaled potential, being dimensionless, 
is expected to become independent of $\mu$. In other words, 
$\hat V_S$ should flow towards an infrared fixed point as 
$\mu\to 0$.

The only true fixed-point solution to 
Eq.~(\ref{eq;RGeqinvhatV0}) is the trivial one, 
$\hat{V}_0=0$. Although there is no other solution that is 
independent of $\mu$ and analytic in $\hat\kappa$, there
is a marginally unstable solution that flows 
logarithmically towards the trivial point. This behavior
is the same as was found in \cite{bb-prc03} using a sharp 
cut-off on the basis of distorted waves. In the current 
scheme, it has the form
\begin{equation}
\frac{1}{\hat{V}_0(\hat\kappa,\mu)}=-1+2\hat\kappa 
\ln\left(\frac{\mu}{\Lambda}\right)\, ,
\label{eq;one_over_hatV0}
\end{equation}
where $\Lambda$ is another arbitrary scale.

We now consider perturbations around this potential that scale
with definite powers of $\mu$. Taking a single term of this form,
\begin{equation}
\frac{1}{\hat{V}_S(\hat{p},\hat{\kappa},\mu)}
= \frac{1}{\hat{V}_0(\hat\kappa,\mu)} 
- C\mu^\nu f(\hat{p},\hat\kappa)\,, 
\end{equation}
and inserting it into the RG equation (\ref{eq;RGeqinvhatV0}),
we find that $f(\hat{p},\hat\kappa)$ satisfies the eigenvalue 
equation
\begin{equation}
\nu f = \hat{p}\,\frac{\partial f}{\partial \hat{p}}
+\hat{\kappa}\,\frac{\partial f}{\partial \hat{\kappa}}
-f \, . 
\end{equation}
Solutions to this are just products of powers of the low-energy 
scales, 
\begin{equation}
f(\hat{p},\hat\kappa)= \hat{p}^{2m}\hat{\kappa}^n\, ,
\end{equation}
and their eigenvalues are
\begin{equation}
\nu=2m+n-1\, .
\end{equation} 
These include one unstable perturbation with a negative 
eigenvalue ($m=n=0$), just as in the 
expansion around the nontrivial fixed point of the pure
short-range potential \cite{bmr-plb99}. There is also a
marginal perturbation with $\nu=0$ ($m=0$, $n=1$) as
expected from the logarithmic dependence on $\mu$ noted
above. In the corresponding power counting, these terms can 
be assigned orders $d=\nu-1=2m+n-2$.

The full short-range potential, expanded around the nontrivial 
solution $\hat V_0$, is given by  
\begin{equation}
\frac{1}{\hat{V}_S(\hat{p},\hat{\kappa},\mu)}
= \frac{1}{\hat{V}_0(\hat\kappa,\mu)} 
- \sum_{m,n\geq 0}C_{mn0}\mu^{2m+n-1} 
\hat{p}^{2m}\hat\kappa^n\,. 
\label{eq;one_over_ED_VS}
\end{equation}
The coefficient of the marginal perturbation, $C_{010}(\Lambda)$ 
depends logarithmically on $\Lambda$ so that the full potential 
is independent of that arbitrary scale.

To interpret the terms in this potential, we solve the LS 
equation with it and calculate the on-shell $T$ matrix. 
Returning to physical units, the potential becomes
\begin{equation}
\frac{1}{V_{S}(p,\kappa,\mu)} = \frac{1}{V_0(\kappa,\mu)} 
- \frac{M}{4\pi}\sum_{m,n\ge 0}C_{mn0}p^{2m}\kappa^n\, ,
\label{eq;one_over_VSE_ph}
\end{equation}
where
\begin{equation}
\frac{1}{V_0(\kappa,\mu)} = \frac{M}{4\pi}\left[
-\mu + 2\kappa\ln\left(\frac{\mu}{\Lambda}\right)\right]\, .
\end{equation}
The corresponding scattering amplitude is
\begin{equation}
\langle \psi^-_C(p)|T_{SC}|\psi^+_C(p)\rangle
= \frac{e^{2i\sigma_0}\,{\cal C}(\kappa/p)}{\displaystyle 
\frac{1}{V_S(p,\kappa,\mu)}-J_0^{\overline{\rm MS}}(p,\kappa,\mu)} \,. 
\label{eq;ED_KS}
\end{equation}
This leads to
\begin{equation}
{\cal C}(\kappa/p)\,p\, \cot\delta_{SC} 
+ 2\kappa \mbox{Re}\{H(\kappa/p)\}
= \sum_{m,n\geq 0}C_{mn0}p^{2m}\kappa^n 
+ 2\kappa \left[1-\gamma
+\ln\left(\frac{\Lambda}{4\kappa}\right)
\right]\, ,
\label{eq;ERE2}
\end{equation}
since the imaginary part of $H$ is
\begin{equation}
\mbox{Im}\{H(\eta)\}=\frac{{\cal C}(\eta)}{2\eta}\, .
\end{equation}

Comparing this with the Coulomb effective-range expansion, 
Eq.~(\ref{eq;ERE}), we see that the coefficients of the 
energy-dependent perturbations are directly related the terms 
in that expansion by
\begin{eqnarray}
-\frac{1}{a_C} &=& C_{000} + C_{010}(\Lambda)\kappa 
+ 2\kappa \left[ 1-\gamma
+\ln\left(\frac{\Lambda}{4\kappa}\right)
\right]+ {\cal O}(\kappa^2)\, ,
\cr
\noalign{\vspace{5pt}}
\frac12\,r_0 &=&
C_{100} + C_{110}\kappa + {\cal O}(\kappa^2)\, .
\end{eqnarray}
If the coefficients in the potential are fine-tuned such
that $1/V_0$ exactly cancels the real part of 
$J_0^{\overline{\rm MS}}$ then the $T$ matrix develops a
pole at $p=0$, corresponding to a zero-energy bound state.
This is the scale-free system, analogous to the one 
corresponding to the fixed point in Ref.~\cite{bmr-plb99}.
In the present case it is unstable against introducing a nonzero 
value for $C_{000}$ or changing $C_{010}$ from the value that 
gives a vanishing coefficient for $\kappa$. 

The full expansion around this point is a double one in powers 
of the energy (or $p^2$) and $\alpha$ (or $\kappa$). For two 
protons $\kappa$ is approximately 3~MeV. This is small compared 
to $m_\pi$ and so this power counting can be used in the context 
of a pionless effective theory. More generally this scale is
given by $\kappa=Z_1Z_2\alpha M_r$, where $M_r$ is the reduced 
mass and the expansion in powers of $\kappa$ can break down if  
it is comparable to the scales of the underlying physics. For 
example, $\kappa$ for the system of two $\alpha$ particles is 
about 60~MeV, and so is of the order of $m_\pi$. In such a 
case, $\kappa$ can be treated as a high-energy scale and the 
resulting pionless effective theory can be expanded in powers of 
energy only, as pointed out by Higa \textit{et al.}~\cite{hhvk-08}.

\section{RG analysis for momentum-dependent potential}

We now turn to more general short-distance potentials that 
depend on momenta as well as energy. For simplicity, we 
consider potentials of the form
\begin{equation}
V_S(p,\kappa,k',k,\mu) = V_1(p,\kappa,\mu) 
+ V_2(p,\kappa,\mu) k^2\, ,
\label{eq;VSk2}
\end{equation} 
where $V_1$ and $V_2$ are energy-dependent functions
and $k$ is the initial relative momentum. 
Although this expression is not Hermitian, it can easily be 
made so by adding a matching term with $k\to k'$.

If we write the corresponding 
(off-shell) $T$ matrix as
\begin{equation}
T_{SC}=T_1+T_2k^2\, ,
\end{equation}
the (regularized) LS equation becomes two coupled linear 
equations
\begin{eqnarray}
T_1&=&V_1+V_1J^{\overline{\rm MS}}_0T_1
+V_2J^{\overline{\rm MS}}_2T_1\,,\cr
\noalign{\vspace{5pt}}
T_2&=&V_2+V_1J^{\overline{\rm MS}}_0T_2
+V_2J^{\overline{\rm MS}}_2T_2\,.
\label{eq;T1T2}
\end{eqnarray}
Here $J_0$ is the bubble integral defined above and
\begin{equation}
J_2=\int \frac{d^3{\bf k}}{(2\pi)^3}\,
\frac{d^3{\bf k}'}{(2\pi)^3}\,
k^2\langle {\bf k}|G_C|{\bf k}'\rangle\,.
\end{equation}

We apply DR and PDS to the integrals over {\bf q} in the same
way for both $J_0$ and $J_2$. However $J_2$ contains an 
additional divergence, in the integral over {\bf k}. This is  
because the Coulomb potential leads to wave functions whose 
derivatives are singular at the origin. We use DR and PDS for
this too but with a different scale, $\lambda$. Using the same 
scale to regulate all the integrals leads to an RG equation 
that contains nonanalytic dependence on the low-energy scales
and is not suitable for a scaling analysis. The second scale
$\lambda$ can be regarded as analogous to the factorization scale
used in parton distributions to separate off the nonperturbative
regime of QCD. Although the factorization scale can, and often 
is, taken to be equal to the renormalization scale, this is not
a requirement. The Wilsonian approach to scattering in 
Ref.~\cite{bb-prc03} used a $\delta$-shell form for the short-range 
potential to avoid the nonperturbative singularities of several 
long-range potentials. The resulting physical amplitudes did not 
depend on the radius chosen for this. In the present case, results 
should be independent of the new scale $\lambda$, and this will 
provide an important consistency check on our treatment. 

With this regularization, we have 
\begin{equation}
J^{\overline{\rm MS}}_2 
= (p^2-2\kappa^2-2\kappa\lambda)J^{\overline{\rm MS}}_0 
- \Delta J_2\, ,
\end{equation}
where $J^{\overline{\rm MS}}_0$ is given by Eq.~(\ref{eq;J0}) and
\begin{equation}
\Delta J_2 = \frac{\pi M}{12}\kappa^2\mu 
+ 2\pi M\kappa^3\zeta'(-2)\,,
\end{equation}
and the derivative of the Riemann zeta function is 
$\zeta'(-2)=-0.0304\cdots$.

By taking appropriate linear combinations of the coupled equations 
for $T_{1,2}$, we can rewrite them as equations that each involve 
only one piece of the potential,\footnote{These are also 
easily obtained from the LS equation in the form 
$T_{SC}=V_S+T_{SC}G_CV_S$, instead of Eq.~(\ref{eq;LSeq}).}
\begin{eqnarray}
T_1&=&V_1\left(1+J^{\overline{\rm MS}}_0T_1
+J^{\overline{\rm MS}}_2T_2\right)\,,\cr
\noalign{\vspace{5pt}}
T_2&=&V_2\left(1+J^{\overline{\rm MS}}_0T_1
+J^{\overline{\rm MS}}_2T_2\right)\,.
\label{eq;dec_LS}
\end{eqnarray}
Demanding that the $T$ matrix be independent of $\mu$, 
$\partial T_{1,2}/\partial \mu$=0, now leads to 
\begin{eqnarray}
0 &=& \frac{\partial V_1}{\partial \mu} \left(
1+J_0^{\overline{\rm MS}}T_1+J_2^{\overline{\rm MS}}T_2\right)
+ V_1\,\frac{\partial J_0^{\overline{\rm MS}}}{\partial\mu}\,
T_1 + V_1\,\frac{\partial J_2^{\overline{\rm MS}}}{\partial\mu}\,T_2\, ,
\cr
\noalign{\vspace{5pt}}
0 &=& \frac{\partial V_2}{\partial \mu} \left(
1+J_0^{\overline{\rm MS}}T_1+J_2^{\overline{\rm MS}}T_2\right)
+ V_2\,\frac{\partial J_0^{\overline{\rm MS}}}{\partial\mu}\,T_1 
+ V_2\,\frac{\partial J_2^{\overline{\rm MS}}}{\partial\mu}\,T_2\, .
\label{eq;dotVs}
\end{eqnarray}
Multiplying these from the right by the factor 
$(1+J_0^{\overline{\rm MS}}T_1+J_2^{\overline{\rm MS}}T_2)^{-1}$ 
and using Eq.~(\ref{eq;dec_LS}) for $T_{1,2}$, we arrive at two coupled 
differential equations for $V_{1,2}$: 
\begin{eqnarray}
\frac{\partial V_1}{\partial\mu}
&=& -V_1^2\,\frac{\partial J^{\overline{\rm MS}}_0}{\partial\mu}
- V_1V_2\,\frac{\partial J^{\overline{\rm MS}}_2}{\partial\mu}\, ,
\cr
\noalign{\vspace{5pt}}
\frac{\partial V_2}{\partial\mu}
&=& -V_1 V_2\,\frac{\partial J^{\overline{\rm MS}}_0}{\partial\mu}
-V_2^2\,\frac{\partial J^{\overline{\rm MS}}_2}{\partial\mu}\, .
\label{eq;dotV2}
\end{eqnarray}

We rescale the potential as before, defining
\begin{equation}
\hat V_1(\hat{p},\hat{\kappa},\mu)
= \frac{M\mu}{4\pi}\, V_1(\mu\hat{p},\mu\hat{\kappa},\mu)\, ,
\qquad
\hat V_2(\hat{p},\hat{\kappa},\mu)
= \frac{M\mu^3}{4\pi}\, V_2(\mu\hat{p},\mu\hat{\kappa},\mu)\, .
\end{equation}
The rescaled functions $\hat V_{1,2}$ then satisfy the coupled RG 
equations
\begin{eqnarray}
\mu\,\frac{\partial}{\partial \mu}\hat{V}_1
&=& \hat{p}\,\frac{\partial}{\partial \hat{p}}\hat{V}_1
+\hat{\kappa}\,\frac{\partial}{\partial \hat{\kappa}}\hat{V}_1
+ \hat{V}_1+ (1-2\hat{\kappa})\hat{V}_1^2\cr 
\noalign{\vspace{5pt}}
&& + \left[(1-2\hat\kappa)\left(\hat{p}^2 -2\hat\kappa^2 
-2\hat\kappa\,\frac{\lambda}{\mu}\right)
+ \frac{\pi^2}{3}\,\hat\kappa^2\right] 
\hat{V}_1\hat{V}_2\, ,
\label{eq;RGeq_for_hatV1}
\\
\noalign{\vspace{5pt}}
\mu\,\frac{\partial}{\partial \mu}\hat{V}_2
&=& \hat{p}\,\frac{\partial}{\partial \hat{p}}\hat{V}_2
+\hat{\kappa}\,\frac{\partial}{\partial \hat{\kappa}}\hat{V}_2
+3 \hat{V}_2
+(1-2\hat\kappa)\hat{V}_1\hat{V}_2\cr
\noalign{\vspace{5pt}}
&&
+\left[(1-2\hat\kappa)\left(\hat{p}^2 -2\hat\kappa^2 
-2\hat\kappa\,\frac{\lambda}{\mu}\right)
+ \frac{\pi^2}{3}\,\hat\kappa^2\right] 
\hat{V}_2^2\, .
\label{eq;RGeq_for_hatV2}
\end{eqnarray}

We are interested in perturbations around the nontrivial solution 
described in the previous section, so we expand the potential 
around $\hat V_0$ as
\begin{equation}
\hat{V}_1 = \hat{V}_0 + \delta\hat{V}_{1}\, ,
\end{equation}
and keep terms to first order in $\delta\hat V_1$ and $\hat V_2$.
The resulting linearized RG equations are
\begin{eqnarray}
\mu\,\frac{\partial}{\partial \mu}\,\delta\hat{V}_{1}
&\simeq& \left[\hat{p}\,\frac{\partial}{\partial \hat{p}}
+\hat{\kappa}\,\frac{\partial}{\partial \hat{\kappa}}\,
+ 1 + 2(1-2\hat\kappa)\hat{V}_0 \right] \delta\hat{V}_{1}\cr
\noalign{\vspace{5pt}}
&& + \left[(1-2\hat\kappa)\left(\hat{p}^2 -2\hat\kappa^2 
-2\hat\kappa\,\frac{\lambda}{\mu}\right)
+ \frac{\pi^2}{3}\,\hat\kappa^2\right] 
\hat{V}_0\hat{V}_2\,,
\label{eq;lin_RGeq_for_hatV1}\\
\noalign{\vspace{5pt}}
\mu\,\frac{\partial}{\partial \mu}\,\hat{V}_2
&\simeq& \left[\hat{p}\,\frac{\partial}{\partial \hat{p}}
+\hat{\kappa}\,\frac{\partial}{\partial \hat{\kappa}}\,
+ 3 + (1-2\hat\kappa)\hat{V}_0 \right] \hat{V}_2\, .
\label{eq;lin_RGeq_for_hatV2}
\end{eqnarray}

The second of these is a homogeneous equation 
for $\hat V_2$ which does not involve $\delta\hat V_1$. By 
comparing it with the equation satisfied by $\hat V_0$, we 
find that it has a solution
\begin{equation}
\hat V_2=C_{001}\mu^2\hat V_0\,,
\end{equation}
with RG eigenvalue $\nu=2$. Other solutions with larger 
eigenvalues can be obtained by multiplying this by powers of
$\hat p^2$ and $\hat\kappa$.

Inserting this solution for $\hat V_2$ into 
Eq.~(\ref{eq;lin_RGeq_for_hatV1}), we get an inhomogeneous
equation for $\delta\hat V_1$. This has the solution
\begin{equation}
\delta\hat V_1=-C_{001}\,\mu^2\left[\left(\hat{p}^2 -2\hat\kappa^2 
-2\hat\kappa\,\frac{\lambda}{\mu}\right)\hat V_0
-\frac{\pi^2}{3}\,\hat\kappa^2\hat V_0^2\right]
+\mu^2\left(A\hat\kappa^3+B\hat\kappa\hat{p}^2\right)\hat V_0^2\, .
\label{eq;deltahatV1}
\end{equation}
The final term here is a solution to the homogeneous part of
the equation, with the same RG eigenvalue as $\hat V_2$. 
It corresponds to two of the terms in Eq.~(\ref{eq;one_over_ED_VS}) 
when that is expanded to first in order in deviations from $\hat V_0$.
The full momentum-dependent perturbation has the form
\begin{equation}
\delta\hat V_S=C_{001}\,\mu^2\left[\left(
\hat k^2-\hat{p}^2+2\hat\kappa^2 
+2\hat\kappa\,\frac{\lambda}{\mu}\right)\hat V_0
+\frac{\pi^2}{3}\,\hat\kappa^2\hat V_0^2\right]  
+\mu^2\left(A\hat\kappa^3+B\hat\kappa\hat{p}^2\right)\hat V_0^2\, .
\end{equation}
As already noted a Hermitian potential can be formed by adding 
the same structure with $k$ replaced by $k'$.

It is worth commenting on several aspects of this term in
the potential. First, having solved the RG equation for fixed 
``factorization scale" $\lambda$, we are now free to choose
$\lambda=\mu$ so that the whole expression is proportional 
to $\mu^2$. The RG eigenvalue is $\nu=2$ and so the term is of 
higher order than the corresponding energy-dependent one, 
$C_{100}\mu\hat{p}^2$ (which has $\nu=1$). This is the same 
pattern as was observed in Ref.~\cite{bmr-plb99} for 
perturbations around the nontrivial fixed point of a pure 
short-range potential. As in that system, the momentum-dependent
perturbation contains the structure $k^2-p^2$ plus terms 
arising from the potential. This form suggests that it 
will vanish when acting on on-shell wave functions and hence
it will alter only the off-shell behavior of the $T$ matrix 
but not observables. We shall see below that this is 
indeed the case.

\section{Scattering by momentum-dependent potential}

We now take the momentum-dependent perturbation constructed in the
previous section and combine it with the energy-dependent potential
of Eq.~(\ref{eq;one_over_VSE_ph}), which we now denote by $V_{SE}$. 
Working in physical units, the potential has the form in 
Eq.~(\ref{eq;VSk2}), with 
\begin{equation}
V_1 = V_{SE} + \delta{V}_1\, ,
\qquad 
V_2 = C_{001}V_0\, ,
\end{equation}
where
\begin{equation}
\delta V_1 = -C_{001} \left[(p^2-2\kappa^2-2\lambda\kappa)V_0 
- \frac{\pi^2}{3}\mu\kappa^2\frac{M}{4\pi}\,V_0^2\right]
+ \left(A\kappa^3+B\kappa p^2\right)\frac{M}{4\pi}\,V_0^2\, .
\end{equation}
Using the LS equation in the form of Eq.~(\ref{eq;T1T2}), we can 
now calculate the $T$ matrix. Expanded to first order in
$\delta V_1$ and $V_2$, this gives
\begin{eqnarray}
T_1 &\simeq&
\frac{1}{1/V_{SE}-J_0^{\overline{\rm MS}}}
+ \left(\frac{1}{1/V_{SE}
-J_0^{\overline{\rm MS}}}\right)^2
\left(
\frac{\delta V_1}{V_{SE}^2} + \frac{V_2}{V_{SE}}J_2^{\overline{\rm MS}}
\right)\, ,\cr
\noalign{\vspace{5pt}}
T_2 &\simeq& \frac{1}{1/V_{SE}-J_0^{\overline{\rm MS}}}\, 
\frac{V_2}{V_{SE}}\, .
\end{eqnarray}

The on-shell $T$ matrix can be written in terms of these as
\begin{equation}
\langle \psi_C^-(p)|T_{SC}|\psi_C^+(p)\rangle = 
\psi_0^2T_1
+ \psi_0\psi_2 T_2 \, , 
\end{equation}
where, following Refs.~\cite{kr-plb99,kr-prc01,ashh-prc07}, 
we have introduced
\begin{eqnarray}
\psi_0 &=& \int\frac{d^3{\bf k}}{(2\pi)^3}\, 
\langle{\bf k} |\psi_C^+(p)\rangle 
= \sqrt{{\cal C}(\kappa/p)}\,e^{i\sigma_0}\, , \cr
\noalign{\vspace{5pt}}
\psi_2 &=& \int\frac{d^3{\bf k}}{(2\pi)^3}\, 
k^2 \langle{\bf k} |\psi_C^+(p)\rangle 
= (p^2-2\kappa^2-2\lambda\kappa)\psi_0\, . 
\end{eqnarray}
Note that we have used DR to regularise the divergent derivative 
of the wave function in $\psi_2$, with the same scale $\lambda$ as 
used above for the similar divergence in $J_2$. Inserting the 
expressions for $T_{1,2}$ from our potential gives
\begin{eqnarray}
&&\frac{e^{-2i\sigma_0}}{{\cal C}(\kappa/p)}
\,\langle \psi_C^-(p)|T_{SC}|\psi_C^+(p)\rangle\cr
&&\qquad \simeq 
\frac{1}{1/V_{SE}-J_0^{\overline{\rm MS}}}
+\left(\frac{1}{1/V_{SE}-J_0^{\overline{\rm MS}}}\right)^2
\left(
\frac{\delta V_1}{V_{SE}^2}
+ \frac{V_2}{V_{SE}^2}\,\frac{\psi_2}{\psi_0}
- \frac{V_2}{V_{SE}}\,\Delta J_2
\right)\, ,
\end{eqnarray}
where we have used the relation 
$J_2\psi_0-J_0\psi_2=-\Delta J_2\psi_0$.

If we now substitute our explicit expressions for $\delta V_1$ 
and $V_2$, we get
\begin{eqnarray}
&&\frac{e^{-2i\sigma_0}}{{\cal C}(\kappa/p)}
\,\langle \psi_C^-(p)|T_{SC}|\psi_C^+(p)\rangle\cr
&&\qquad \simeq 
\frac{1}{1/V_{SE}-J_0^{\overline{\rm MS}}}
+\left(\frac{1}{1/V_{SE}-J_0^{\overline{\rm MS}}}\right)^2
\frac{{V}_0}{{V}_{SE}}
\left[
C_{001}\frac{\pi M}{12} \mu \kappa^2
\left(\frac{{V}_0}{{V}_{SE}}-1\right)\right.\cr
\noalign{\vspace{5pt}}
&&\hspace{9cm} -2\pi C_{001} M \kappa^3\zeta'(-2)\cr
\noalign{\vspace{5pt}}
&& \hspace{9cm}
\left.+ \frac{M}{4\pi}\left(A\kappa^3+B\kappa p^2\right)
\frac{V_0}{V_{SE}}\right]\, .
\label{eq;TSC}
\end{eqnarray}
In solving the RG we kept only terms to first order in deviations
from the fixed point. For consistency we should make the same 
approximation here, replacing $V_{SE}$ by $V_0$ in the second term
of this expression. The first term in the square bracket then vanishes.
The remaining terms in cancel if we make the choices,
\begin{equation}
A = 8\pi^2 C_{001} \zeta'(-2)\, ,\qquad B=0,
\end{equation}
for the constants of integration in our momentum-dependent perturbation.
The on-shell scattering amplitude then reduces to the same form, 
Eq.~(\ref{eq;one_over_VSE_ph}), as we obtained for the purely 
energy-dependent potential. This shows that the momentum-dependent
eigenfunction of the RG equation is indeed an ``equation of motion"
term, affecting only the off-shell behaviour of the $T$ matrix.

\section{Discussion}

We have applied the RG to $S$-wave scattering by a combination of 
Coulomb and short-range potentials, looking in particular at strongly 
interacting systems with bound states close to threshold. Unlike
previous work \cite{kr-plb99,kr-prc01,bb-prc03,ashh-prc07}, our 
approach treats both energy- and momentum-dependence of the 
short-range potential separately. To handle the divergences generated 
by the contact interactions we use dimensional regularization 
with the PDS and $\overline{\rm MS}$ schemes to subtract the 
linear and logarithmic divergences. The RG equation is obtained 
by requiring that the off-shell scattering matrix be independent of
the subtraction scale.

The resulting RG equation has a physically interesting nontrivial 
solution. This is not an exact fixed point because it possesses a
marginally relevant perturbation and so it evolves logarithmically 
with the regulator scale. Perturbations around this can be expanded 
in powers of energy, momenta and $\alpha$ (or the corresponding 
low-energy scale $\kappa=\alpha M/2$). 
The RG eigenvalues and hence
the power counting for purely energy-dependent perturbations are the
same as those found for systems with only short-range interactions
\cite{bmr-plb99}. This reflects the fact that the Coulomb and
free wave functions have the same power-law behavior near the 
origin. The fact that the Coulomb potential leads to a logarithmic
discontinuity in derivatives of the wave functions is reflected 
by the presence of the marginal term proportional to $\kappa$.

Like the fixed point of the pure short-range potential, the 
nontrivial solution describes a system with a bound state at zero
energy. The double expansion around it in powers of the energy and 
$\kappa$ corresponds to the Coulomb effective-range expansion
originally introduced by Bethe \cite{b-pr49}, with the coefficients 
expanded in powers of $\alpha$.

We have extended this approach to terms that depend on the off-shell 
momenta, presenting in detail the analysis for the simplest of these,
which is second-order in the momenta. The results for the power
counting match with those found in the pure short-range case 
\cite{bmr-plb99}. (Again, this  is not surprising given the behavior 
of the Coulomb wave functions at the origin.) In particular, the 
momentum-dependent terms appear at higher orders than the 
corresponding energy-dependent ones. The one studied here has an RG 
eigenvalue $\nu=2$, compared with $\nu=1$ for the term proportional 
to the on-shell energy.

The structure of these momentum-dependent terms indicates that 
they arise from the equation of motion and hence they should 
affect only the off-shell behavior of the scattering matrix. We 
have verified that our second-order term does not contribute to 
scattering observables. 

This RG analysis can explain some of the puzzling features seen in 
other applications of EFTs to charged-particle scattering.
The different power counting for energy- and momentum-dependent
perturbations around the nontrivial fixed point
means that one cannot ``use the equations of motion"
to remove energy-dependence from the effective potential without 
mixing terms of different orders. Requiring the potential to be 
energy-independent for all values of the regulator scale is only
possible if the higher-order momentum dependent perturbations are 
given unnaturally large coefficients. This is quite different from
expansions around a trivial fixed point where the corresponding 
terms appear at the same order. 

Most approaches to the EFT for charged-particle scattering are 
based on purely momentum-dependent potentials (derivatives of 
$\delta$-functions)~\cite{kr-plb99,kr-prc01,ashh-prc07}. Each term 
in such a potential can be built out of a combination of 
energy-dependent perturbations and off-shell ones. The 
leading-order piece of each is the energy-dependent one and so 
these potentials can reproduce the Coulomb effective-range expansion. 
However, as mixtures of different RG eigenfunctions, these terms 
contain pieces that violate the power counting. As a result, 
the inclusion of higher-order terms can significantly change the 
coefficients of lower-order ones. 

Moreover, the same on-shell perturbation can form part of more 
than one distinct momentum-dependent term. For example, 
Ref.~\cite{ashh-prc07} shows that both the structures 
$k^4+k^{\prime 4}$ and $k^2k^{\prime 2}$ can generate a shape
parameter in the effective-range expansion. However they contain
different admixtures of off-shell perturbations and hence they 
affect the renormalization of the lower-order terms differently,
as demonstrated by the numerical results in that paper.

Finally, the combination of momentum-dependent perturbations 
and the Coulomb potential leads to new divergences. These arise
from derivatives of the Coulomb wave functions at the origin.
We have used DR and PDS to regularize these but we found it 
necessary to keep their subtraction scale distinct from the one 
used for the integral over intermediate states. We regard
this second scale as analogous to a factorization scale, 
separating off the nonperturbative regime caused by the
singularity in the long-range interaction at the origin. 
Although we have been unable to obtain a useful RG 
equation by demanding that the off-shell scattering matrix 
be independent of this scale, we are able to show that 
physical observables do not depend on it.

\section*{Acknowledgments}

This work was supported by STFC grant number PP/F000448/1.


\vskip 3mm \noindent
\begin{thebibliography}{99}
\bibitem{betal-00}
S.~R. Beane {\it et al.}, 
in \textit{At the Frontier of Particle Physics: Handbook of QCD},
edited by M. Shifman (World Scientific, Singapore, 2001), 
Vol.~1, p.~133.

\bibitem{bk-arnps02}
P.~F. Bedaque and U. van Kolck,
Ann.~Rev.~Nucl.~Part.~Sci. {\bf 52}, 339 (2002).
\bibitem{e-ppnp06}
E. Epelbaum,
Prog.~Part.~Nucl.~Phys. {\bf 57}, 654 (2006)
\bibitem{b-pr49}
H.~A. Bethe, Phys.~Rev. {\bf 76}, 38 (1949).

\bibitem{kr-plb99}X. Kong and F. Ravndal,
Phys. Lett. \textbf{B450}, 320 (1999).
\bibitem{kr-prc01} X. Kong and F. Ravndal,
Phys. Rev. C \textbf{64}, 044002 (2001).
\bibitem{ashh-prc07} S. Ando, J.~W. Shin, C.~H. Hyun and 
S.-W. Hong, Phys. Rev. C \textbf{76}, 064001 (2007).

\bibitem{bb-prc03} T. Barford and M.~C. Birse,
Phys. Rev. C \textbf{67}, 064006 (2003).
\bibitem{wk-pr74}
K.~G. Wilson and J.~G. Kogut,
Phys.~Rep. {\bf 12}, 75 (1974).
\bibitem{bmr-plb99} M.~C. Birse, J.~A. McGovern and 
K.~R. Richardson, Phys. Lett. \textbf{B464}, 169 (1999). 

\bibitem{ksw-plb98}
D.~B. Kaplan, M.~J. Savage, M.~B. Wise,
Phys.~Lett. {\bf B 424}, 390 (1998).
\bibitem{ksw-npb98}
D.~B. Kaplan, M.~J. Savage, M.~B. Wise,
Nucl.~Phys. {\bf B 534}, 329 (1998).

\bibitem{hhvk-08}
R. Higa, H.-W. Hammer and U. van Kolck, 
arXiv:0802.3426. 

\end{thebibliography}
\end{document}